\documentclass[english,aps,floats,twocolumn,showpacs,nofootinbib]{revtex4}
\usepackage{pslatex}
\usepackage[T1]{fontenc}
\usepackage[latin1]{inputenc}
\usepackage{graphicx}
\usepackage{epsfig}

\usepackage{calc}
\usepackage{ifthen}

{
{
{
\newcommand{\bea}{\begin{eqnarray}}
\newcommand{\eea}{\end{eqnarray}}

\newcommand{\nc}{\newcommand}
\nc{\renc}{\renewcommand}
\nc{\eqs}[2]{\mbox{Eqs.~(\ref{#1},\,\ref{#2})}}
\nc{\eq}[1]{\mbox{Eq.~(\ref{#1})}}
\nc{\figs}[2]{\mbox{Figs.~(\ref{#1},\,\ref{#2})}}
\nc{\fig}[1]{\mbox{Fig~.(\ref{#1})}}
\nc{\be}[1]{\begin{equation} \mbox{$\label{#1}$}}
\nc{\ee}{\vspace{0.1cm}\end{equation}}

\newcommand{\bean}{\begin{eqnarray*}}
\newcommand{\eean}{\end{eqnarray*}}

\def\GeV{{\rm \ GeV}}

\def\lae{\;^{<}_{\sim} \;}

\begin{document}
\title{Explaining the Dark Energy, Baryon and Dark Matter Coincidence via Domain-Dependent Random Densities}
\author{John McDonald}
\email{j.mcdonald@lancaster.ac.uk}
\affiliation{Lancaster-Manchester-Sheffield Consortium for Fundamental Physics, Cosmology and Astroparticle Physics Group, Dept. of Physics, University of 
Lancaster, Lancaster LA1 4YB, UK}
\begin{abstract}

    The dark energy, dark matter and baryon densities in the Universe are observed to be similar, with a factor of no more than 20 between the largest and smallest densities. We show that this coincidence can be understood via superhorizon domains of randomly varying densities when the baryon density at initial collapse of galaxy-forming perturbations is determined by anthropic selection. The baryon and dark matter densities are assumed to be dependent on random variables $\theta_{d}$ and $\theta_{b}$ according to $\rho_{dm} \propto \theta_{d}^{\alpha}$ and $\rho_{b} \propto \theta_{b}^{\beta}$, while the effectively constant dark energy density is dependent upon a random variable $\phi_{Q}$ according to $\rho_{Q} \propto \phi_{Q}^{n}$.  The ratio of the baryon density to the dark energy density at initial collapse, $r_{Q}$, and the baryon-to-dark matter ratio, $r$, are then determined purely statistically, with no dependence on the anthropically-preferred baryon density. We compute the probability distribution for $r_{Q}$ and $r$ and show that the observed values of $r_{Q}$ and $r$ can be naturally understood within this framework.   In particular, for the case $\alpha = 2$, $\beta = 1$ and $n = 4$, which can be physically realized via a combination of axion dark matter, Affleck-Dine baryogenesis and frozen quintessence with a $\phi_{Q}^4$ potential, the range of $r_{Q}$ and $r$ which corresponds to the observed Universe is a quite natural, with a probability which is broadly similar to other ranges of $r_{Q}$ and $r$.

\end{abstract}
\maketitle

\section{Introduction} 

     A fundamental issue for cosmology is the similarity of the observed densities of dark energy, dark matter and baryons. There have been several proposals which attempt to explain either the ratio of the dark energy density to the total matter density or the ratio of the baryon density to the dark matter density. Combining these proposals can, in principle, account for the observed coincidence of all three densities. 

    It has been suggested that the dark energy density could be explained by anthropic selection \cite{wein,wein2}. This is because density perturbation growth is damped in a dark energy dominated Universe. Therefore if there is a variation of the dark energy density between domains of size much larger than the presently observed Universe, we might be able to understand why the dark energy density is close to the total matter density today.

      There have also been many proposals for explaining the baryon-to-dark matter ratio. A particularly popular approach is based on a conserved charge shared between baryons and dark matter, which results in similar number densities of dark matter particles and baryons. In most cases this requires that the dark matter be asymmetric, with mass in the range 1-10 GeV. 

     If dark matter is not related to the baryon number by a conserved charge, then we need to consider a different explanation for the baryon-to-dark matter ratio. One way is to have similar physical processes which independently determine the baryon and dark matter densities. For example, right-handed sneutrino dark matter can be generated by a process similar to Affleck-Dine leptogenesis \cite{snu}. This approach allows a wider range of dark matter particle mass, since there is no direct link between the number densities of the dark matter particles and the baryons. A particular class of such models arises if dark matter is a thermal WIMP. In this case the observed baryon density must be related to thermal WIMP annihilation \cite{bm1}. There have been a number of proposals to relate the baryon density to the thermal "WIMP miracle" density \cite{bm1,bm2,wimpy,baryowimp,davidson}, either by modifying a pre-existing baryon asymmetry \cite{bm1,bm2} or by generating the baryon asymmetry from thermal WIMP annihilations or decays \cite{wimpy,baryowimp,bernal}. 

  Combining any of these proposals for the baryon-to-dark matter ratio with anthropic selection of the cosmological constant could then explain the coincidence of all the observed densities.

    Alternatively, we could consider the possibility that all three densities have a domain dependence. For example, a domain-dependent dark matter density is natural in the case of axion dark matter \cite{lindeax}, while a domain-dependent baryon asymmetry is possible in Affleck-Dine baryogenesis \cite{lindead,aadb}. In fact, there may be a strong motivation for considering a domain-dependent baryon asymmetry. An increase in the average baryon density at a fixed temperature will cause a much larger increase in the baryon density in galaxies \cite{lindeax,aadb}. As a result, the observed baryon asymmetry is likely to be rather close to the anthropic upper bound on the baryon asymmetry. To understand this additional coincidence may then require that the baryon asymmetry is anthropically selected via domain dependence. The baryon-to-dark matter ratio could then be either due to a conserved charge or due to domain dependence of the dark matter density. We will consider the latter possibility in the following.  

   In \cite{domain1} we presented a general analysis of the baryon-to-dark matter ratio in the case where both the baryon and dark matter densities have a domain dependence due to random variables which correspond to effectively massless fields during inflation. The key assumption of the model is that the baryon density when the perturbations responsible for galaxies initially become non-linear and begin to collapse is the dominant anthropic parameter, such that equal values of the baryon density at initial collapse correspond to equal probability for the evolution of observers in galaxies. In this case the baryon-to-dark matter ratio $r$ is purely statistical and is completely independent of the baryon density at initial collapse \cite{domain1}. It was shown that observed baryon-to-dark matter ratio, $r \approx 1/5$, is then a completely natural value in this framework. 

    Our previous discussion did not include dark energy. Here we consider whether random domains of all three densities can naturally account for the observed coincidence. To do this we must compute the probability distribution of the dark energy, dark matter and baryon densities for a given dependence on the random variables. 

  In the following we will analyse the probability distribution in a completely general way, without specifying the origin of the random variables. The simplest possibility is that the random variables are fields which are effectively massless during inflation and are randomized by quantum fluctuations \cite{aadb,domain1}.  We will show that domain dependence of the baryon, dark matter and dark energy densities, when combined with the principle that the baryon density at initial collapse is the primary anthropic selection parameter, can naturally account for the observed coincidence of all three densities. 

   As an explicit example of a domain-dependent dark energy density, we will consider a frozen quintessence model \cite{cald} with a $\phi_{Q}^{n}$ potential, where the mass and couplings are sufficiently small that the quintessence field $\phi_{Q}$ is effectively frozen at the present epoch. When combined with axion dark matter and Affleck-Dine baryogenesis, this can provide an explicit realization of the class of models we are considering here. 

   The paper is organized as follows. In Section 2 we derive the general probability distribution for the ratio of the baryon density to the dark energy density at initial collapse, $r_{Q}$, and the baryon-to-dark matter ratio, $r$, for galaxies with equal baryon density at initial collapse.
We also derive the constraints on possible models from the requirement that the probability distribution is well-defined. In Section 3 we apply this distribution to a range of models 
and show that values of $r_{Q}$ and $r$ which are consistent with the observed Universe are quite natural in a number of models. In Section 4 we present  our conclusions. In the Appendix we discuss frozen quintessence models and show that they can consistently generate a domain-dependent frozen dark energy density via quantum fluctuations.

\section{Probability Distribution for $r_{Q}$ and $r$}

     In \cite{domain1} we considered dark matter and baryon densities which depend on random variables $\theta_{d}$ and $\theta_{b}$ according to $\rho_{d} \propto \theta_{d}^{\alpha}$ and $\rho_{b} \propto \theta_{b}^{\beta}$. In the model of \cite{domain1}, $\theta_{d}$ and $\theta_{b}$ are the phases of massless fields which take random values during inflation due to quantum fluctuations, such that all values of the angles are equally probable\footnote{We should make clear what we mean by a "domain". Quantum fluctuations randomize the fields at a number of e-foldings much larger than $N \approx 60$, when the observed Universe exits the horizon. At $N \approx 60$, there will then be approximately constant random values of the fields in any horizon-sized volume at $N \approx 60$, due to the random walk sum of fluctuations of wavelength much larger than the horizon at $N \approx 60$ \cite{aadb}. There will also be isocurvature fluctuations of wavelength close to the horizon at $N \approx 60$, but these are small perturbations of the otherwise constant value of the random fields over the horizon.}. However, we may consider $\theta_{d}$ and $\theta_{b}$  to be general random variables. The anthropic selection condition that galaxies with equal baryon density at initial collapse are anthropically equally probable independently of $r$ was then applied. Although an idealization, this condition is likely to be a good approximation for the range of baryon-to-dark matter ratio favoured by the resulting probability distribution, $0.1 \lae r \lae 10$ \cite{domain1}. It should be emphasized that the only quantities which vary between the domains are the densities. In particular, the primordial density perturbation is assumed to be the same in all domains.

     Under these condition, the probability density function $f_{r}(r)$ of the baryon-to-dark matter ratio, $r$, was shown to be determined purely by the statistical probability of the corresponding values of $\theta_{d}$ and $\theta_{b}$ \cite{domain1}, 
\be{d7}  f_{r}(r) = N \left(\frac{r}{1 + r}\right)^{\frac{3}{4}\left(\frac{1}{\alpha} + \frac{1}{\beta}\right)} 
\frac{1}{r^{\frac{1}{\alpha} + 1}}  ~,\ee
where $N$ is a normalization constant which depends on $\alpha$ and $\beta$. Rather remarkably, this is completely independent of the specific value of the baryon density at initial collapse, $\rho_{b\;g}$. 

   Physically, this distribution arises because different values of $r$ can result in the same baryon density at initial collapse of galaxy perturbations. For example, if we increase the average dark matter density at a fixed temperature, the effect will be to increase the temperature at matter-radiation equality, resulting in earlier perturbation growth and so earlier initial collapse at a higher temperature. This increases the baryon density in galaxies at initial collapse. This can be compensated by reducing the average baryon density at a fixed temperature, so keeping the baryon density at initial collapse fixed. The probability of $r$ is then determined by the number of domains of different average dark matter and baryon density which can produce a given baryon density at initial collapse.

     We now consider the effect of a domain-dependent dark energy density. We will consider a constant dark energy density $\rho_{Q}$ which depends on a random variable $\phi_{Q}$ according to $\rho_{Q} 
\propto \phi_{Q}^{n}$, where all values of $\phi_{Q}$ are considered equally probable. In the frozen quintessence models discussed in the Appendix, $\phi_{Q}$ is the quintessence field and its potential is $V(\phi_{Q}) \propto \phi_{Q}^{n}$. 

   The effect of increasing the dark energy density will be to damp perturbation growth once the total matter density is less than the dark energy density. Therefore initial collapse of the galaxy-forming perturbations must occur before the total matter density $\rho_{TOT}$ is less than the dark energy density. In the following we will consider a simplified model where the growth of perturbations stops once $\rho_{Q} > \rho_{TOT}$ and is unaffected when $\rho_{Q} < \rho_{TOT}$.

   We next compute the probability for a domain to have a values of $(r_{Q}, r)$ in the range $(r_{Q \; 1},r_{1})$ to $(r_{Q\;2}, r_{2})$. Here $r_{Q} = \rho_{b\;g}/\rho_{Q}$ is the ratio of the baryon density at initial collapse to the constant dark energy density. 

    There are two cases. If $r_{Q} > 1$ then $\rho_{TOT} > \rho_{Q}$ for all $r$ at initial collapse. Therefore the dark energy density has no effect on galaxy formation. The probability density $f$ of the domains is then simply the product of the probability density $f_{r}$ of $r$ and the probability density $f_{r_{Q}}$ of $r_{Q}$, where the factors are independent
\be{p1} f(r_{Q},r) = f_{r_{Q}}(r_{Q})f_{r}(r)    ~.\ee
The probability density for $r_{Q}$ at a fixed value of $\rho_{b\;g}$ is obtained from the flat distribution for $\phi_{Q}$ (i.e. all $\phi_{Q}$ having equal probability), 
\be{p1a}  {\rm Prob.} \propto d\phi_{Q} \propto \frac{dr_{Q}}{r_{Q}^{
\frac{n + 1}{n}}} \Rightarrow f_{r_{Q}} \propto   \frac{1}{r_{Q}^{
\frac{n + 1}{n}}} ~.\ee
On the other hand, if $r_{Q} < 1$ then, since $\rho_{b\;g} < \rho_{Q}$, we require 
a sufficiently large dark matter density in order to have $\rho_{TOT} > \rho_{Q}$ at initial collapse. This imposes an upper bound, $r_{c}$, on $r$ as a function of $r_{Q}$,
\be{p2}  r < r_{c}(r_{Q}) = \frac{r_{Q}}{1 - r_{Q}}    ~,\ee
where $r_{c} = \infty$ once $r_{Q} \geq 1$. 
This in turn reduces the total number of domains available at a given $r_{Q}$.

    The probability $P$ to be in the range $r_{Q\;1}$ to $r_{Q\;2}$ 
and $r_{1}$ to $r_{2}$ is then 
\be{p4} P = N \int_{r_{Q\;1}}^{r_{Q\;2}} \int_{r_{1}}^{Min(r_{2},r_{c}(r_{Q}))} \frac{1}{r_{Q}^{\frac{n + 1}{n}}}  \left(\frac{r}{1 + r}\right)^{\frac{3}{4}\left(\frac{1}{\alpha} + \frac{1}{\beta}\right)} 
\frac{1}{r^{\frac{1}{\alpha} + 1}}  \; dr \; dr_{Q}   ~,\ee 
where the overall normalization $N$ satisfies  
\be{p4a} N \int_{0}^{\infty} \int_{0}^{r_{c}(r_{Q})} \frac{1}{r_{Q}^{\frac{n + 1}{n}}}  \left(\frac{r}{1 + r}\right)^{\frac{3}{4}\left(\frac{1}{\alpha} + \frac{1}{\beta}\right)} 
\frac{1}{r^{\frac{1}{\alpha} + 1}} \; dr \; dr_{Q} = 1  ~.\ee 
$P$ simply counts the total number of available domains with 
values of $r$ and $r_{Q}$ within the given ranges, subject to the conditions that $\rho_{b\;g}$ is constant and $\rho_{TOT} > \rho_{Q}$ at initial collapse. An important feature is that the probability distribution for $r_{Q}$ and $r$ is independent of the specific value of the baryon density at initial collapse $\rho_{b \;g}$, which is assumed to be determined by anthropic selection. 

   \eq{p4} is a well-defined probability as long as the integral in \eq{p4a} is finite. As shown in 
\cite{domain1}, the integral over $r$ is finite provided that $\alpha > \beta/3$. The integral over $r_{Q}$ can diverge at small $r_{Q}$. In the limit where the upper limit on the $r_{Q}$ integral is $r_{Q\;u} \ll 1$, $r_{c}(r_{Q})$ is approximately equal to $r_{Q}$  and the integral becomes 
\be{p6}  N \int_{0}^{r_{Q\;u}} \frac{1}{r_{Q}^{\frac{n + 1}{n}}}  \int_{0}^{r_{Q}}  r^{\frac{3}{4}\left(\frac{1}{\alpha} + \frac{1}{\beta}\right)} 
\frac{1}{r^{\frac{1}{\alpha} + 1}}  \; dr \; dr_{Q} =   N \int_{0}^{r_{Q\;u}} r_{Q}^{\frac{3}{4\beta} - \frac{1}{4\alpha} -\frac{(n+1)}{n} } dr_{Q}  ~.\ee 
Thus the condition for convergence as the upper limit $r_{Q\;u} \rightarrow 0$ is 
\be{p7} \frac{3}{4\beta} - \frac{1}{4\alpha} -\frac{(n+1)}{n} + 1 > 0 ~.\ee
If the integral does not converge then the probability distribution will strongly prefer small values of $r_{Q}$ and $r$, corresponding to very large dark energy and dark matter densities relative to the baryon density. We expect that the dark matter density will be anthropically cut-off when it becomes sufficiently large, because the increase in the dark matter density in the galaxy halo for a given baryon density will cause the radius of the disk to become smaller for a given disk angular momentum. In this case, assuming that the effect of increasing the 
dark matter density is dominated by the increase in mass of the halo (and not by an increase in the angular momentum of the disk), the density of stars in the disk will eventually become large enough to affect stable solar system formation. However, we expect this to become important only at very small values of $r$ and therefore not to alter the conclusion that very large dark matter and dark energy densities will be favoured in this case. Therefore it is unlikely that models with divergent probability distributions will be consistent with the observed Universe. This is a significant constraint on possible models.

\section{Probability of the density ratios in specific models}

    In this section we will consider some plausible examples of models, which are defined by $\alpha$, $\beta$ and $n$. We will consider models with $n = 2$ and $n = 4$. These can be realized by frozen quintessence models with $\phi_{Q}^2$ and $\phi_{Q}^4$ potentials respectively (Appendix). For $\alpha$ and $\beta$ we will consider models with $(\alpha, \beta)$ = (1,1), (2,1), (1,2) and (2,2). $(\alpha,\beta) = (2,1)$ can be realized by axion dark matter with an approximately $a^2$ potential combined with a baryon asymmetry 
determined by a domain-dependent CP-violating phase, as in Affleck-Dine baryogenesis \cite{aadb}. $(\alpha, \beta) = (1, 1)$ could be realized by asymmetric dark matter proportional to a domain-dependent CP-violating phase combined with a baryon asymmetry which is also proportional to a domain-dependent CP-violating phase.

   The range relevant to our domain is $r_{Q} = 100 - 1000$ and $r = 0.1-1$. The range of $r$ is determined by the magnitude of the observed baryon-to-dark matter ratio, $r = 0.2$. To estimate the range of $r_{Q}$ which is consistent with the observed Universe, we need to estimate the dark energy density when the galaxy perturbations first go non-linear and begin to collapse. The red-shift at the initial collapse of a galaxy perturbation, $z_{coll}$, is related to $r_{Q}$ and the present density of baryons and dark energy by 
\be{p5} r_{Q}  = \left( 1 + z_{coll}\right)^{3}\frac{\Omega_{b}}{\Omega_{Q}}       ~.\ee 
With $\Omega_{b} = 0.04$ and $\Omega_{Q} = 0.73$ we find that $r_{Q} \approx 95$ if $z_{coll} \approx 11$ and $r_{Q} \approx 1000$ if $z_{coll} \approx 25$. Therefore $r_{Q}$ in our domain is in the range 100-1000 if $z_{coll}$ is in the range $11-25$. This range is consistent with estimates of the time of formation of proto-galaxies.

   We first consider which models have convergent probability distributions. For $n = 2$ the convergence condition is 
\be{n1} \frac{3}{\beta} - \frac{1}{\alpha}  > 2   ~.\ee  
This can only be satisfied by $(\alpha, \beta) = (2,1)$. 
For $n = 4$ the convergence condition is 
\be{n2} \frac{3}{\beta} - \frac{1}{\alpha}  > 1   ~.\ee
This can be satisfied by $(\alpha, \beta) = (2,1)$ and (1, 1). Models with $(\alpha, \beta) = (1, 2)$ and $(2, 2)$ are generally ruled out by convergence when $n = 2$ or 4.
 
   We next discuss the probability distributions for the three convergent models and their compatibility with the observed Universe.

  We first consider the case $n = 2$ and $(\alpha, \beta) = (2,1)$. In Table 1 we give the probability for $r_{Q}$ and $r$ to be within different ranges of values, where we express the probability in terms of the percentage of the total number of domains. This shows that the probability of $r_{Q}$ and $r$ being within the observed range in this model is small, with only 0.9$\%$ of domains having $r_{Q} = 100-1000$ and $r = 0.1-1$. This can be compared with the most likely range, $r_{Q} = 0.1 - 1$ and $r = 0.1-1$, which has 13.5$\%$ of the domains and is therefore 14.8 times more likely. Thus domains with dark energy densities which are much larger than those we observe are strongly preferred in this model.
This is also apparent from Figure 1, which shows the strong preference for domains with much smaller $r_{Q}$ (i.e. larger dark energy density relative to baryons) than we observe. The prospects for the observed value of $r$ are better. Figure 2 shows that values of $r$ in the range 0.1-1 are generally the most likely or close to the most likely when $r_{Q} > 0.1$.

    Consistency with the observed Universe is much improved in the model with $n = 4$ and $(\alpha,\beta) = (2,1)$. From Table 2 we see that the probabilities for all ranges of $r_{Q}$ and $r$ are not very much different. The probability (12.1$\%$) of the most likely range of $r_{Q}$ and $r$ ($r_{Q} = 1-10$ and $r = 1-10$), is only 3.5 times larger than the probability (3.4$\%$) of range corresponding to our domain ($r_{Q} = 100-1000$ and $r = 0.1-1$). This is also clear from Figures 3 and 4. From Figure 3 we see that the probability distribution of $r_{Q}$ is more evenly spread across the ranges of $r_{Q}$ than in the case $n = 2$, with the peak of the distribution being at larger values of $r_{Q}$. Therefore values of $r_{Q}$ in the range consistent with the observed Universe are a natural possibility in this model, with a probability which is not very much smaller than the other ranges of $r_{Q}$. From Figure 4 we see that the probability of $r$ in the range 0.1-1.0 is large for the most likely ranges of $r_{Q}$. 
Therefore the model with $n = 4$ and $(\alpha,\beta) = (2,1)$ can naturally account for the observed coincidence of the dark energy, dark matter and baryon densities. 

    The case $n = 4$ and $(\alpha, \beta) = (1,1)$ is similar to $n = 4$ and $(\alpha, \beta) = (2,1)$. From Table 3 and Figures 5 and 6 we see that the distribution of values of $r_{Q}$ is similar to, but slightly less strongly peaked than, the case  $(\alpha, \beta) = (2,1)$. The probability of $r_{Q}$ and $r$ being in the range consistent with our observed Universe is again relatively large (3.6$\%$) compared with the most likely range ($r_{Q} = 0.1-1$ and $r = 0.1-1$), which is 3.1 times more likely with 11.1$\%$ of the domains. Therefore it is again quite reasonable in this model to find ourselves in a domain with the observed dark energy, dark matter and baryon densities.     

   The preference of the large value of $r_{Q}$ observed in our domain for models with larger values of $n$ can be easily understood. The gain in probability in going to a larger value of $\rho_{Q}$  is smaller in this case, since 
a given increase in $\phi_{Q}$ produces a larger increase 
in the dark energy density $\rho_{Q} \propto \phi_{Q}^{n}$. 
Therefore larger dark energy densities (corresponding to smaller $r_{Q}$) are less strongly favoured.

\begin{table}[h]
\begin{center}
\begin{tabular}{|c|c|c|c|c|c|}
 \hline $r_{Q}$	&	$r = 0.001-0.1$     &   $0.01-0.1$  &  $0.1-1$ & $1-10$ & $10-100$ \\
\hline	$0.001-0.01 $	&	$9.8$ & $0$	&	$0$	&	$0$	& $0$ \\   	
\hline	$0.01-0.1 $	&	$8.0$ & $9.6$	&	$0.03$	&	$0$	& $0$ \\   	
\hline	$0.1-1$	&	$2.5$ & $9.7$	& $13.5$	&	$2.8$	& $0.15$	\\ 
\hline	$1-10$	& $0.8$ &	$3.1$      & $8.9$	&	$10.2$	& $4.6$	\\   		
\hline	$10-100$	&	$0.26$ & $1.0$ & $2.9$	&	$3.2$	& $1.4$ 	\\   	
\hline	$100-1000$	&	$0.08$ & $0.31$ &  $0.91$	&	$1.0$	 & $0.46$	\\   		
\hline	$1000-10000$	&	$0.04$ & $0.11$ &  $0.33$	&	$0.37$	 & $0.17$	\\   
\hline
 \end{tabular} 
 \caption{\footnotesize{Percentage of domains within ranges of  $(r_{Q}, r)$  for $n = 2$ and $(\alpha,\beta) = (2,1)$.} } 
 \end{center}
 \end{table}

\begin{table}[h]
\begin{center}
\begin{tabular}{|c|c|c|c|c|c|}
 \hline $r_{Q}$	&	$r = 0.001-0.01$ & $0.01-0.1$  &  $0.1-1$ & $1-10$ & $10-100$\\
\hline	$0.001-0.01 $	&	$2.3$ & $0$	&	$0$	&	$0$	& $0$ \\ 

\hline	$0.01-0.1 $	&	$3.0$ & $4.2$	&	$0.02$	&	$0$	& $0$ \\ 
\hline	$0.1-1 $	&	$1.7$ & $6.4$	& $10.0 $	&	$4.0$	  & $0.14$	\\   
\hline	$1-10 $	&	$0.95$ & $3.6$    & $10.6	$   & $12.1$	& $5.3$	\\   		
\hline	$10-100 $	&	$0.54$ & $2.1$	& $6.1$	&	$6.9$	  & $3.0$ 	\\   	
\hline	$100-1000 $	&	$0.30$ & $1.2$ &  $3.4$	&	$3.9$	 & $1.7$	\\   		
\hline	$1000-10000$	&	$0.20$ & $0.77$ &  $2.3$	&	$2.6$	 & $1.2$	\\   		
\hline
 \end{tabular} 
 \caption{\footnotesize{Percentage of domains within ranges of  $(r_{Q}, r)$  for $n = 4$ and $(\alpha,\beta) = (2,1)$.}  }
 \end{center}
 \end{table}

\begin{table}[h]
\begin{center}
\begin{tabular}{|c|c|c|c|c|c|}
 \hline $r_{Q}$	&	$r = 0.001-0.01$ & $0.01-0.1$  &  $0.1-1$ & $1-10$ & $10-100$\\
\hline	$0.001-0.01 $	&	$5.2$	&	$0$ & $0$	&	$0$	& $0$ \\ 
\hline	$0.01-0.1 $	&	$6.4$ & $6.9$	&	$0$	&	$0$	& $0$ \\ 
\hline	$0.1-1 $	&	$3.6$ & $9.9$	& $11.1 $	&	$1.6$	  & $0$	\\   
\hline	$1-10 $	&	$2.0$ & $5.5$    & $11.1$   & $6.7$	& $1.1$	\\   		
\hline	$10-100 $	&	$1.1$ & $3.2$	 & $6.3$	&	$3.8$	  & $0.65$ 	\\   	
\hline	$100-1000 $	&	$0.65$ & $1.8$ &  $3.6$	&	$2.2$	 & $0.36$	\\   		
\hline	$1000-10000$	&	$0.43$ & $1.2$ &  $2.4$	&	$1.4$	 & $0.24$	\\   		\hline
 \end{tabular} 
 \caption{\footnotesize{Percentage of domains within ranges of $(r_{Q}, r)$  for $n = 4$ and $(\alpha,\beta) = (1,1)$.} }  
 \end{center}
 \end{table}

\begin{figure}[htbp]
\begin{center}
\epsfig{file=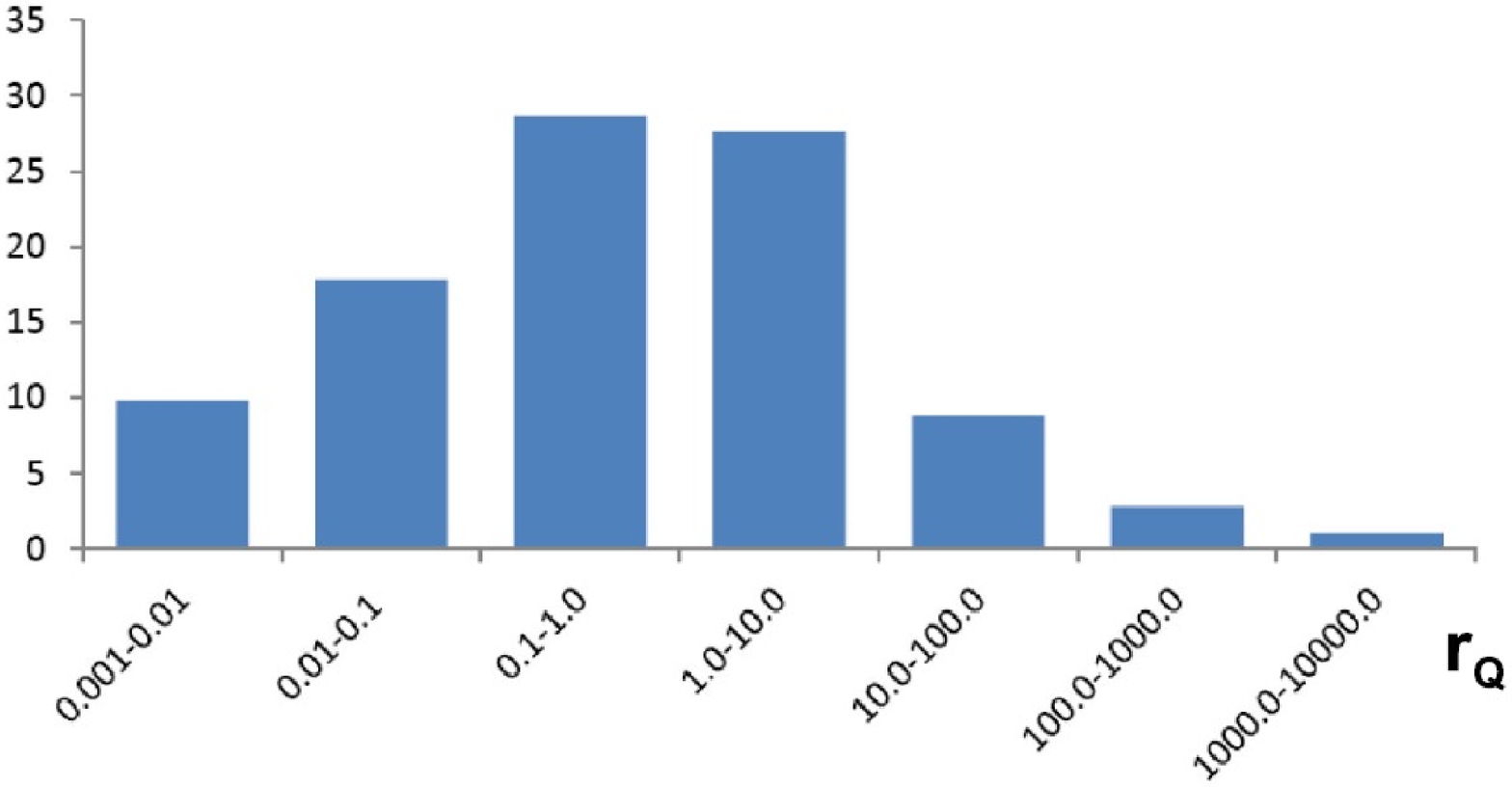, width=0.5\textwidth, angle = 0}
\caption{Percentage of domains as a function of the baryon-to-dark energy ratio at initial collapse $r_{Q}$ for $n = 2$ and $(\alpha,\beta) = (2,1)$}
\label{fig2}
\end{center}
\end{figure}

\begin{figure}[htbp]
\begin{center}
\epsfig{file=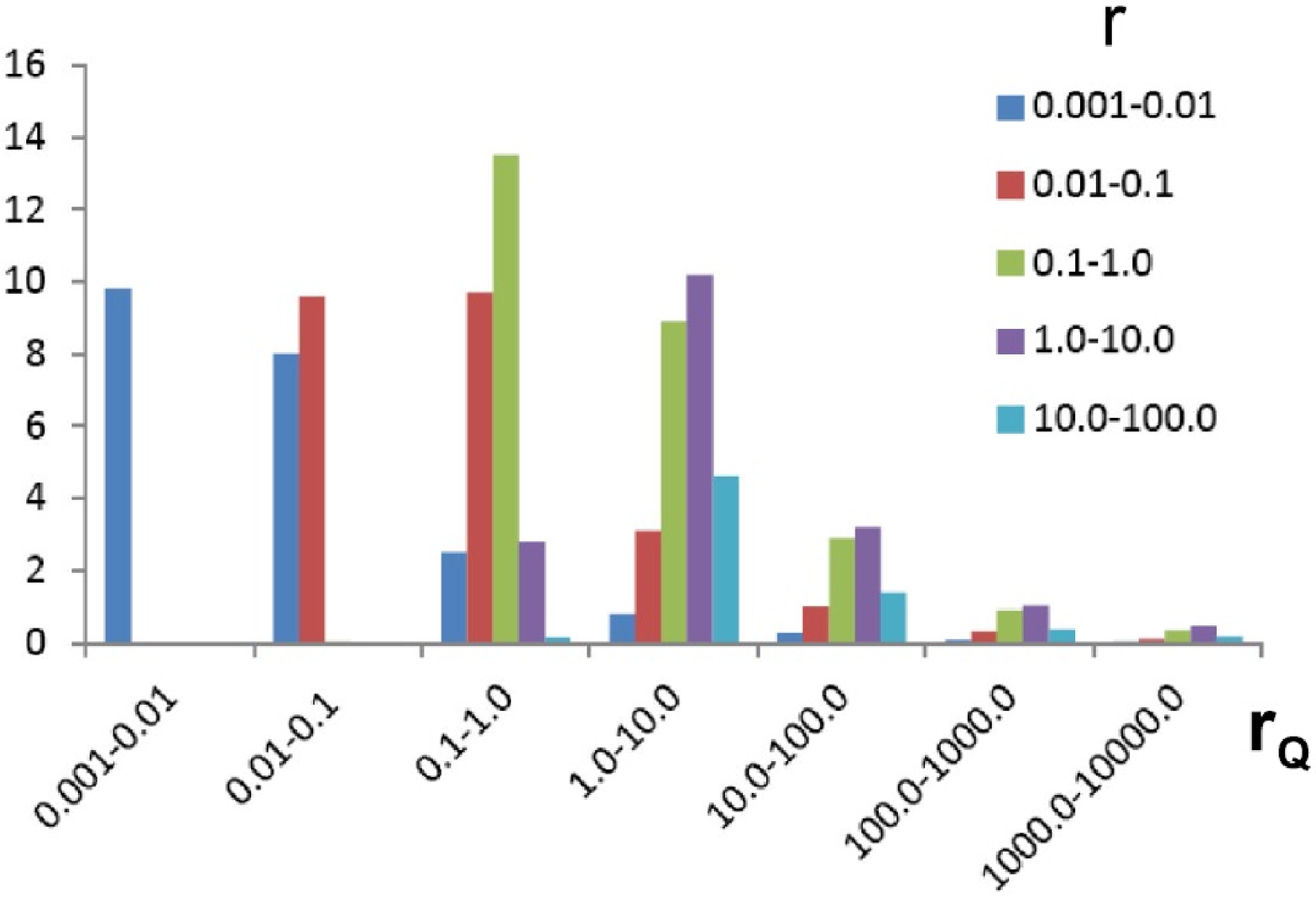, width=0.5\textwidth, angle = 0}
\caption{Percentage of domains as a function of the baryon-to-dark energy ratio at initial collapse $r_{Q}$ and the baryon-to-dark matter ratio $r$ for $n = 2$ and $(\alpha,\beta) = (2,1)$}
\label{fig1}
\end{center}
\end{figure}

\begin{figure}[htbp]
\begin{center}
\epsfig{file=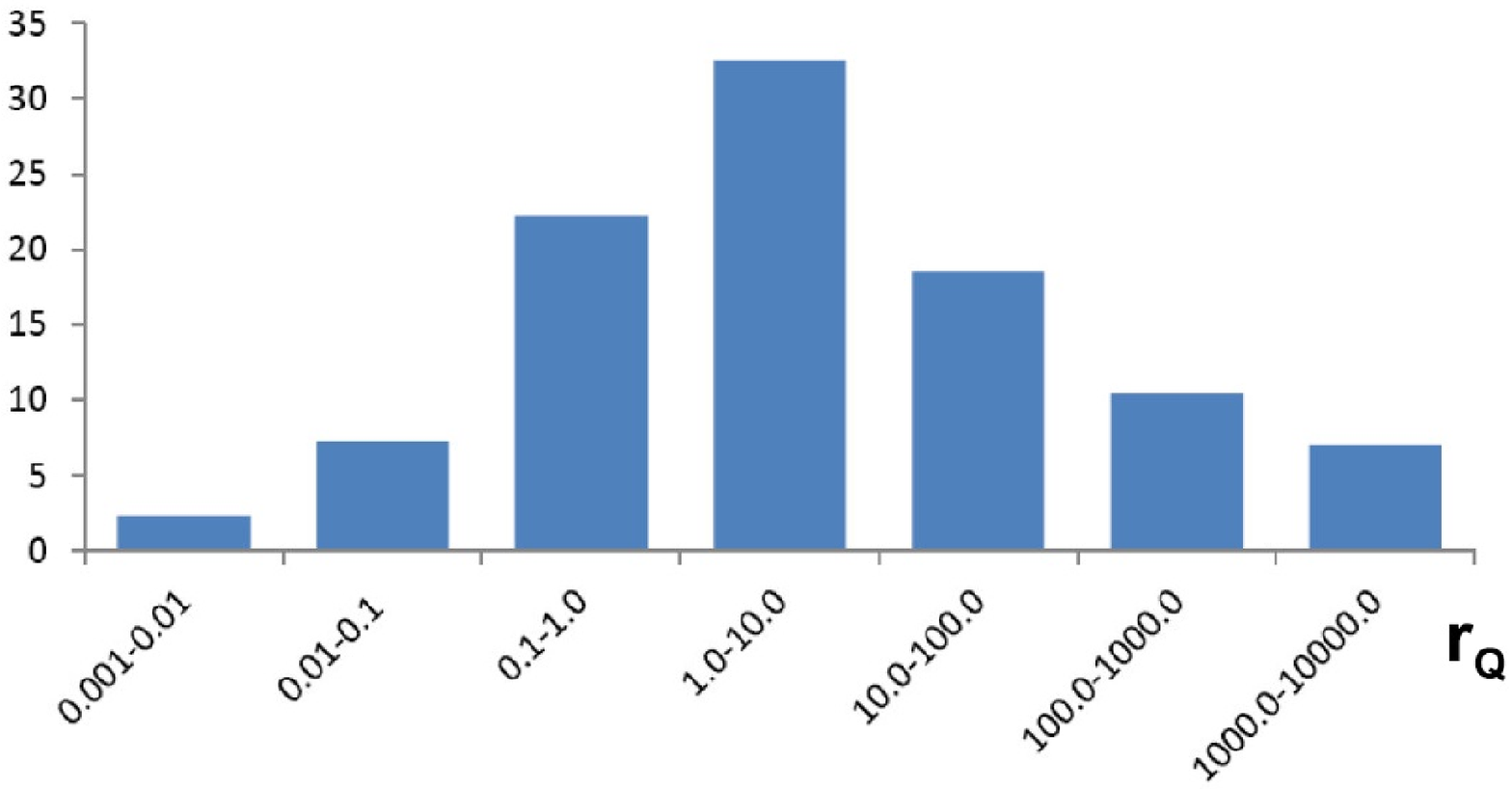, width=0.5\textwidth, angle = 0}
\caption{Percentage of domains as a function of the baryon-to-dark energy ratio at initial collapse $r_{Q}$ for $n = 4$ and $(\alpha,\beta) = (2,1)$}
\label{fig4}
\end{center}
\end{figure}

\begin{figure}[htbp]
\begin{center}
\epsfig{file=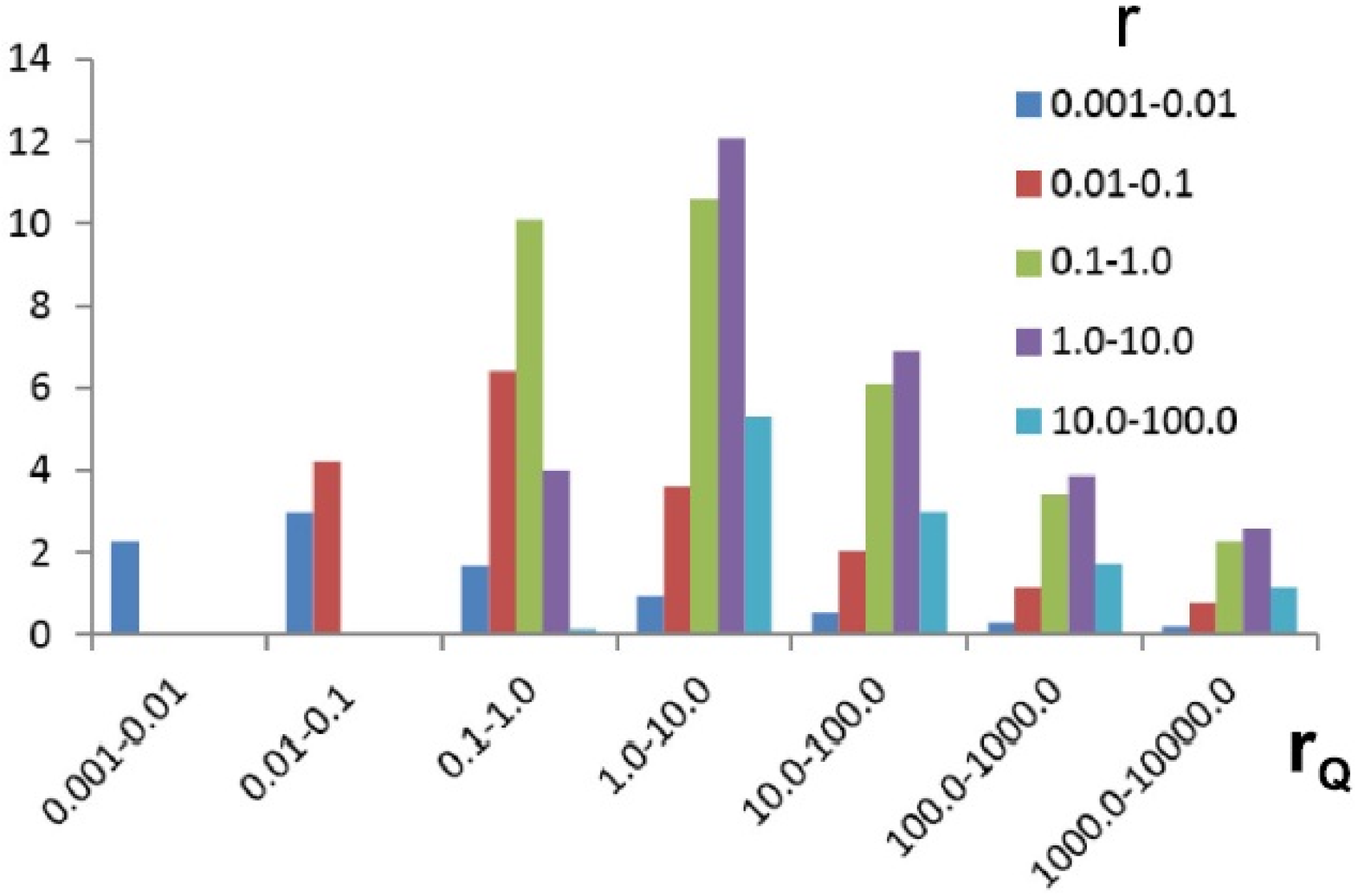, width=0.5\textwidth, angle = 0}
\caption{Percentage of domains as a function of the baryon-to-dark energy ratio at initial collapse $r_{Q}$ and the baryon-to-dark matter ratio $r$ for $n = 4$ and $(\alpha,\beta) = (2,1)$}
\label{fig3}
\end{center}
\end{figure}

\begin{figure}[htbp]
\begin{center}
\epsfig{file=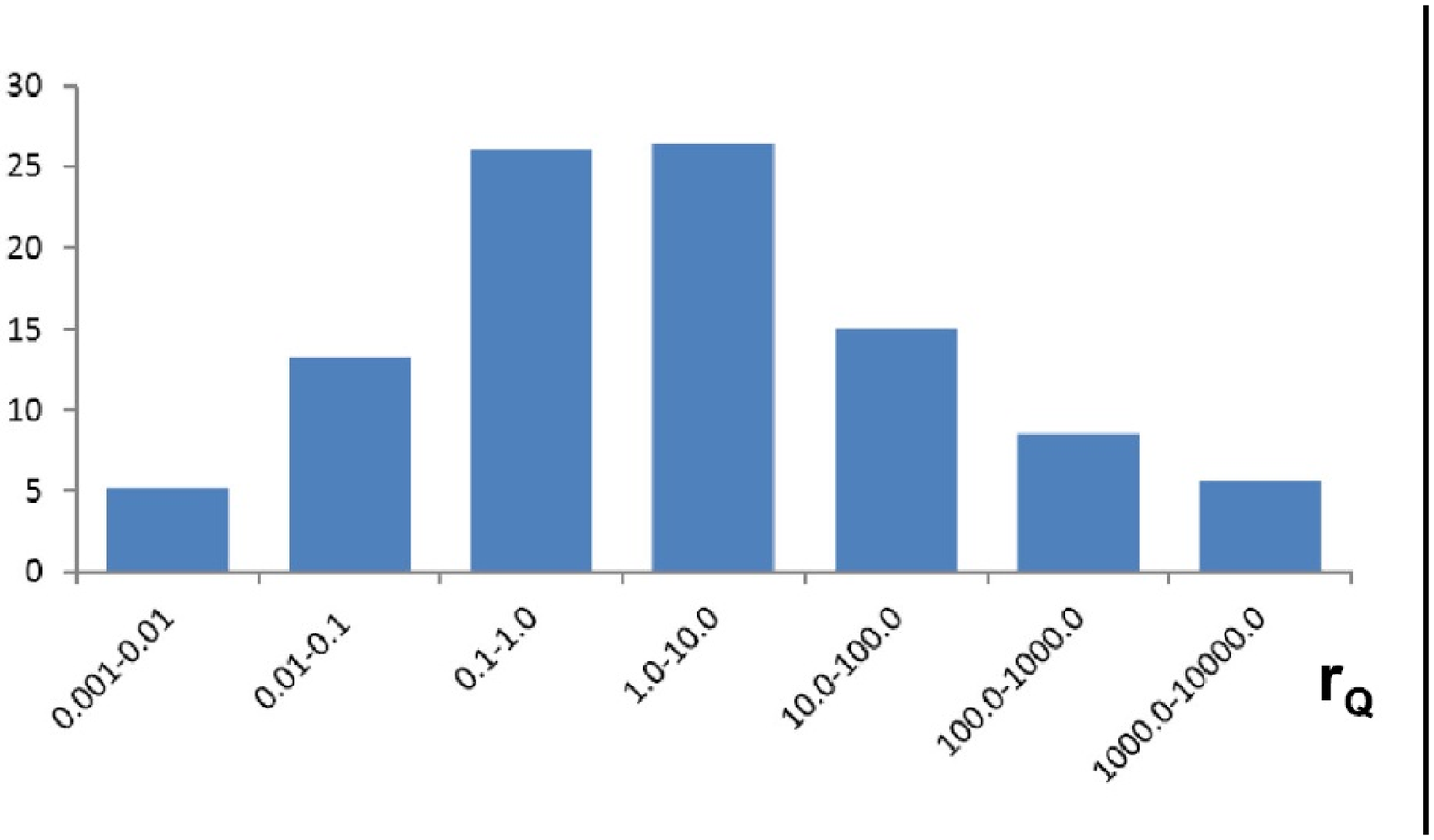, width=0.5\textwidth, angle = 0}
\caption{Percentage of domains as a function of the baryon-to-dark energy ratio at initial collapse $r_{Q}$ for $n = 4$ and $(\alpha,\beta) = (1,1)$}
\label{fig6}
\end{center}
\end{figure}

\begin{figure}[htbp]
\begin{center}
\epsfig{file=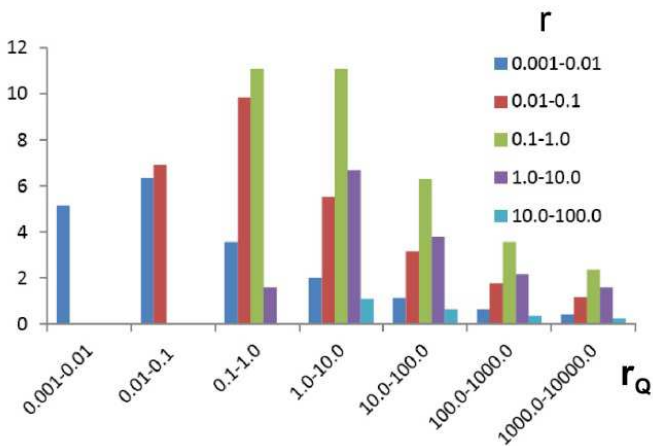, width=0.5\textwidth, angle = 0}
\caption{Percentage of domains as a function of the baryon-to-dark energy ratio at initial collapse $r_{Q}$ and the baryon-to-dark matter ratio $r$ for $n = 4$ and $(\alpha,\beta) = (1,1)$}
\label{fig5}
\end{center}
\end{figure}

\section{Conclusions}

       We have shown that is possible to explain the coincidence of the observed dark energy, dark matter and baryon densities via domains of varying densities generated by random variables combined with anthropic selection which is primarily sensitive to the baryon density at the initial collapse of galaxy-forming perturbations.

      We have derived the complete probability distribution for the number of domains with fixed baryon density at initial collapse as a function of the baryon-to-dark energy ratio at initial collapse, $r_{Q}$, the baryon-to-dark matter ratio, $r$, and the model parameters $n$, $\alpha$ and $\beta$, where the respective densities depend on the random variables $\phi_{Q}$, $\theta_{d}$ and $\theta_{b}$ according to $\rho_{Q} \propto \phi_{Q}^{n}$, $\rho_{dm} \propto \theta_{d}^{\alpha}$ and  $\rho_{b} \propto \theta_{b}^{\beta}$. 
This allows us to compute the probability of finding ourselves in a domain with a given magnitude of $r_{Q}$ and $r$. The probability distribution for $r_{Q}$ and $r$ is essentially statistical in nature, being independent of the anthropically-selected baryon density. A natural interpretation of the random variables is that they are effectively massless fields during inflation, whose domain dependence is generated by quantum fluctuations. 

   An important constraint on possible models is the requirement that the total probability converges to a finite value. We find that for $n = 2$, which could be realized by a $\phi_{Q}^2$ frozen quintessence model, this is only possible for models with $(\alpha,\beta) = (2,1)$, while for $n = 4$, which could be realized by a $\phi_{Q}^4$ frozen quintessence model, it is possible for  models with $(\alpha,\beta) = (2,1)$ and (1,1). $(\alpha,\beta) = (2,1)$ could be realized by, for example, axion dark matter combined with Affleck-Dine baryogenesis (or, more generally, any form of baryogenesis based on a domain-dependent CP-violating phase), while $(\alpha, \beta) = (1,1)$ 
could be realized by asymmetries of both dark matter and baryons which are determined by domain-dependent CP-violating phases. 

   Comparing the probability for $r_{Q}$ and $r$ with the ranges consistent with the observed Universe, we find that the model with $n = 2$ is not favoured, with only a small fraction of the domains being consistent with the observed Universe, making our domain atypical. On the other hand, we find that in models with $n = 4$, our domain is a more natural possibility. The distribution of dark energy densities has broadly similar probabilities over a wide range of $r_{Q}$, while for larger values of $r_{Q}$ the most likely value of $r$ lies in the range 0.1-10. The domains with $r_{Q}$ and $r$ in the range consistent with the observed Universe are fairly typical, with only about 3.5 times smaller probability than the most likely range of values of $r_{Q}$ and $r$. Thus it is quite natural to find values of the dark matter and dark energy densities relative to the baryon density which are consistent with those we observe, even though larger values of the dark energy density are more probable. 

   The fundamental assumption of the model is that anthropic selection is primarily sensitive to the baryon density at initial collapse and is relatively insensitive to the value of $r$. For baryon densities at initial collapse similar to the observed Universe, anthropic sensitivity to $r$ is likely to become important only at very large dark matter densities, corresponding to small values of $r \ll 0.1$. These values of $r$ make only a small contribution to the probability distribution for $r$, therefore the model should be reliable in this case. Thus it should be possible to explain the dark matter and dark energy densities in our domain via domain-dependent densities.  Nevertheless, it would be interesting to consider the effect on the final state of galaxies of varying the baryon density and baryon-to-dark matter ratio at initial collapse. This would allow us to assess the anthropic selection of the baryon density and the likelihood of the observed baryon density, and to quantify the anthropic sensitivity to $r$. This is quite challenging, as we would need to understand the effect of different initial baryon and dark matter densities on the typical size of galaxies, the process of gas cloud collapse and star formation and the final state of the galactic disks.

    The framework we have discussed here can account for the similarity of the present density of dark energy to that of dark matter and baryons only if the dark energy model can generate domains of extremely small dark energy density with a suitable probability distribution. Therefore the problem of the dark energy coincidence is effectively replaced by the problem of the specific nature of dark energy. Only once this is definitively understood can we claim to have a complete explanation of the coincidence.

     There exist other dark energy models which can generate a probability distribution for the dark energy density. In particular, the landscape scenario proposes that a very large number of meta-stable vacuum states are nucleated in an eternal inflation-like evolution of the Universe. This effectively generates domains in which all possible vacuum states are realized with a probability distribution. The implications of this scenario for the cosmological constant were discussed in \cite{bousso}.  This approach (the "causal patch measure" \cite{bousso}) is largely independent of the specific nature of the observers, in particular whether they require galaxies. It depends only on the number of possible vacuum states plus some broad assumptions about the probability of the existence of observers at a given time. It is therefore unrelated to the approach we have discussed here. However, the landscape can also play a role in the framework we consider here, which is specific to observers that require galaxies. Provided that the probability distribution for the cosmological constant from the landscape favours larger values of $\Lambda$ and leads to a convergent probability distribution for the density ratios, the landscape can serve as a source of the dark energy density in our framework. The baryon and dark matter domains could then either arise via field theory mechanisms within a specific dark energy domain or as part of the landscape itself.

   In conclusion, we find that it is possible to naturally account for the coincidence of dark energy, dark matter and baryons in a framework in which all densities are determined by random variables and the dominant anthropic selection parameter is the baryon density in galaxies at the initial collapse of galaxy-forming perturbations.

\section*{Acknowledgements}
The work of JM is supported by the Lancaster-Manchester-Sheffield Consortium for Fundamental Physics under STFC grant
ST/J000418/1.

\renewcommand{\theequation}{A-\arabic{equation}}
 \setcounter{equation}{0} 

\section*{Appendix: Frozen quintessence as a model for a domain-dependent dark energy density}

 As an example of a model of a domain-dependent dark energy density, we will consider a simple frozen quintessence model based on a potential $V(\phi_{Q}) = K \phi_{Q}^n$ , where the effective mass squared $V^{''}(\phi_{Q})$ of $\phi_{Q}$ is much less that the present value of the Hubble parameter squared, $H_{o}^{2}$. (Although $V^{''}(\phi_{Q})$ could be close to $H_{o}^{2}$, this would amount to an unlikely coincidence, therefore we will consider $V^{''}(\phi_{Q}) \ll H_{o}^{2}$ to be the natural condition.) The domains will then be generated by quantum fluctuations of the effectively massless quintessence field. As a background inflation model we will consider $\phi^2$ chaotic inflation. 

  The condition $V^{''}(\phi_{Q}) \ll H_{o}^2$ at present imposes an upper bound on the value of $K$, 
\be{a1} K  \ll K_{max} = \frac{H_{o}^{2}}{n(n-1)\phi_{Q}^{n-2}}       ~.\ee
The condition for the energy density to be dominated by dark energy at present is that $H_{o}^2 = V(\phi_{Q})/3 M_{p}^{2}$. ($M_{p}$ is the reduced Planck mass throughout, $2.4 \times 10^{18} \GeV$.) Therefore 
\be{a2}   \phi_{Q} = \left( \frac{3 M_{p}^2 H_{o}^2}{K} \right)^{1/n}   ~.\ee 
Combining \eq{a1} and \eq{a2} implies that 
\be{a3} \phi_{Q} \gg (3n (n-1))^{1/2} M_{p}   ~.\ee
Therefore $\phi_{Q} \gg M_p$ is necessary to have frozen quintessence. 

  We next show that $\phi_{Q}$ must be generated by quantum fluctuations in order to satisfy $\phi_{Q} \gg M_{p}$. We assume that $\phi_{Q}$ is effectively massless during inflation. We also assume that the length scale of initial fluctuations of the fields is characterised by the initial chaotic inflation horizon. In this case the initial fluctuations $\delta \phi_{Q}$ cannot be larger than $M_p$. This follows from the contribution of the gradient term to the energy density
\be{e1} \rho_{Q} \sim (\nabla \phi_{Q})^2 \sim H^2 \delta \phi_{Q}^2   ~.\ee
Therefore $\rho_{Q} \lae \rho = 3 M_p^2 H^2$ implies that $\delta \phi_Q \lae M_p$. Therefore if we wish to have $\phi_{Q} \gg M_p$ in domains at $N = 60$, we require that quantum fluctuations of $\phi_Q$ can subsequently drive the field to large values. 

     We next show that this can easily be achieved in $\phi^2$ chaotic inflation. We consider the initial condition to be $H \approx M_{p}$, corresponding to a Planck-scale initial energy density, as expected from chaotic initial conditions. The initial energy density of the inflating patch is $V(\phi_{i})$, 
where $\phi_{i}$ is the initial value of the inflaton field.   
Therefore
\be{e2} \phi_{i} \approx \frac{\sqrt{6} M_{p}^{2}}{m_{\phi}}   ~.\ee
The number of e-foldings until the end of inflation $N$ for a given $\phi$ is 
\be{e3}   N = 2 \pi G (\phi^2 -\phi_{end}^2) \approx 2 \pi G \phi^2   ~,\ee
where $\phi_{end} \approx \sqrt{2} M_{p} (\ll \phi_{i})$ from the condition $\eta(\phi_{end})  \approx 1$. 
Therefore there is a very large total number of e-foldings from $\phi  = \phi_{i}$ to $\phi_{end}$, 
\be{e4}  N_{TOT} \approx 2 \pi G \phi_{i}^2 \approx \frac{3}{2} \left( \frac{M_{p}}{m_{\phi}} \right)^2 ~.\ee
$m_{\phi} \approx 1.5 \times 10^{13} \GeV$ is necessary to generate the correct magnitude of primordial density perturbation, therefore $N_{TOT} \approx 4 \times 10^{10}$. 

   The average quantum fluctuation of the quintessence field on the scale of the horizon per e-folding is $\delta \phi_{Q} = H/2 \pi$. If $H$ is constant, the resulting random walk of the field will result in a Gaussian distribution for the 
field in a given horizon volume. 
In fact, $H$ is not constant during $\phi^2$ chaotic inflation, 
since $H = (2 m_{\phi}^2 N/3)^{1/2}$ at $N$ e-foldings. Therefore the probability distribution for the effectively constant value of $\phi_{Q}$ in a domain at $N \approx 60$ will not be exactly Gaussian, since the random walk step of the $\phi_{Q}$ field is time-dependent. Nevertheless, we can compute the variance $\sigma$ of the 
field by summing the variance due to small numbers of e-foldings, since $H$ varies significantly only on a time scale much larger than an e-folding. Therefore 
\be{e4a} \sigma^2 = <\delta \phi_{Q}^2> = \int_{0}^{N_{TOT}} \frac{H^{2}(N)}{4 \pi^2} dN = \frac{H^{2}(N_{TOT})}{4 \pi^2} \times \frac{N_{TOT}}{2}   ~.\ee
With $N_{TOT} \approx 4 \times 10^{10}$ we obtain $\sigma \approx  2 \times 10^{4} M_{p}$.

     Thus quantum fluctuations of the quintessence field can generate superhorizon-sized domains at $N = 60$ with effectively constant values of $\phi_{Q}$ which are much larger than $M_{p}$.      
For values $\phi_{Q} \ll \sigma$, all values of $\phi_{Q}$ will be equally probable, as in the case of an exact Gaussian distribution. This can be seen since most of the variance is generated at $N > N_{TOT}/2$, where $H$ is approximately constant and the distribution of $\phi_{Q}$ is close to Gaussian.     

     $\phi_{Q}$ at present can remain constant at all values up to the variance $\sigma$ if $V^{''}(\sigma) \ll H_{o}^2$. This is true if 
\be{a4} K \ll \frac{H_{o}^2}{n(n-1) \sigma^{n-2}}    ~.\ee
For $n = 2$ this requires that $m_{Q} \ll H_{o}$, since 
$K = m_{Q}^{2}/2$. Therefore $m_{Q} \ll 10^{-42} \GeV$ is necessary.  For $n = 4$ the condition becomes 
\be{a5} K < \frac{2 \pi^2 H_{o}^{2}}{3 N_{TOT} M_{p}^2} \approx 10^{-130}   ~.\ee      
Therefore a severe tuning of the $\phi_{Q}^4$ self-coupling is necessary to allow $\phi_{Q}$ to take any value out to the variance. The value of $\phi_{Q}$ in our domain could be much smaller than the variance, in which case larger $K$ could be considered. However, even with the smallest possible value of $\phi_{Q}$, $\phi_{Q} \approx M_{p}$, the $n=4$ model still requires $K < 10^{-120}$. We note that a sufficiently large maximum value of $\phi_{Q}$ compared with $M_{p}$ is necessary to allow a large enough range of dark energy density for anthropic selection to function. 

    A combination of super-Planckian $\phi_{Q}$ and extremely small mass and self-coupling terms for $\phi_{Q}$ is  therefore necessary to have a successful frozen quintessence model with dark energy domains generated by quantum fluctuations. This appears to be a rather extreme set of assumptions. However,  the assumptions of the frozen quintessence model are simply a more extreme version of the assumptions behind the $\phi^2$ chaotic inflation model. In the $\phi^2$ chaotic inflation model $\phi(N) \approx 2 \sqrt{N} M_{p} > M_{p}$ at $N$ e-foldings and $m_{\phi}/M_{p} \approx 10^{-5}$. Therefore the difference between the tuning of the two models is only a question of degree, which will depend on the measure for the parameters (the difference in tuning appears less extreme if viewed logarithmically).

    Finally, it is important to check that quantum fluctuations of $\phi_{Q}$ generated at $N \approx 60$ do not lead to dangerous fluctuations of the dark energy on observable scales. If we consider a fluctuation $\delta \phi_{Q}$ then the energy density in the fluctuation once within the horizon is 
\be{a6} \delta \rho_{Q} \approx V^{'}(\phi_{Q}) \delta \phi_{Q}    ~,\ee
where we assume that the energy of the fluctuation is dominated by its potential, which is true if ${\bf k}^{2} \delta \phi_{Q}^{2} < V^{'}(\phi_{Q}) \delta \phi_{Q}$, where ${\bf k}$ is the wavenumber of the fluctuation. If we consider a fluctuation entering the horizon at the present time, corresponding to a perturbation which exited the horizon at $N \approx 60$, then $\delta \phi_{Q} \approx H(N = 60)/2 \pi$. With $V(\phi_{Q}) \propto \phi_{Q}^{n}$, we find 
\be{a7} \frac{ \delta \rho_{Q}}{\rho_{Q}} = \frac{n \delta \phi_{Q}}{\phi_{Q}} = \frac{n H(N = 60)}{2 \pi \phi_{Q}}  ~.\ee
Thus 
\be{a8} \frac{ \delta \rho_{Q}}{\rho_{Q}} \sim \frac{m_{\phi} \sqrt{N}}{M_{p}}\times \frac{M_{p}}{\phi_{Q}} \approx 5 \times 10^{-5} \frac{M_{p}}{\phi_{Q}}    ~.\ee
Since typically $\phi_{Q} \gg M_{p}$,  it follows that $\delta \rho_{Q}/\rho_{Q} \ll 10^{-5}$ and so quantum fluctuations of the quintessence field will have a completely negligible effect. The condition  ${\bf k}^{2} \delta \phi_{Q}^{2} < 
V^{'}(\phi_{Q}) \delta \phi_{Q}$ is satisfied in the case 
${\bf k}^2 \sim H_{o}^{2}$ if $H(N = 60) < M_{p}^{2}/\phi_{Q}$, which is generally true for the models considered here.  

    Thus frozen quintessence with a $\phi_{Q}^{n}$ potential can consistently generate domains of different dark energy density. The probability distribution of $\phi_{Q}$ is flat, so $\phi_{Q}$ acts as a random variable which generates an effective cosmological constant in each superhorizon-sized domain at present. This provides an existence proof for such models, although, as with all quintessence models, rather extreme assumptions the regarding mass and coupling terms are necessary.


\end{document}